\documentclass[sigconf, nonacm]{acmart}
\AtBeginDocument{%
  \providecommand\BibTeX{{%
    \normalfont B\kern-0.5em{\scshape i\kern-0.25em b}\kern-0.8em\TeX}}}

\usepackage{natbib}
\usepackage{caption,subcaption,tabularx}
\usepackage{graphicx,float}
\usepackage[utf8]{inputenc}
\usepackage{enumitem}
\usepackage{amsmath}
\usepackage{todonotes}

\begin{document}

\title{Patent Search Using Triplet Networks Based Fine-Tuned SciBERT}

%% used to denote shared contribution to the research. Be
\author{Utku Umur Acikalin}
\email{u.acikalin@etu.edu.tr}
\orcid{1234-5678-9012}

\affiliation{%
  \institution{TOBB University of Economics and Technology}
  \city{Ankara}
  \country{Turkey}
}

\author{Mucahid Kutlu}
\email{m.kutlu@etu.edu.tr}
\orcid{1234-5678-9012}

\affiliation{%
  \institution{TOBB University of Economics and Technology}
  \city{Ankara}
  \country{Turkey}
}

\renewcommand{\shortauthors}{Acikalin and Kutlu}

\begin{abstract}
In this paper, we propose a novel method for the prior-art search task. We fine-tune SciBERT  transformer model  using Triplet Network approach, allowing us to represent each patent with a fixed-size vector. This also enables us to conduct efficient vector similarity computations to rank patents in query time. In our experiments, we show that our proposed method outperforms  baseline methods. 
\end{abstract}






\maketitle

\section{Introduction}

%\setlist{noitemsep}  % Reduce space between list items (itemize, enumerate, etc.)
%\onehalfspacing      % Use 1.5 spacing

%Patents are vital for Intellectual Property protection and encourage people to work on innovative research and development. 
The number of patents is increasing rapidly with the incredible advances in scientific knowledge and technology. This brings many challenges for patent examiners as they have to compare each patent application against prior ones and determine whether it is novel. Therefore, we need effective search engines that can find relevant patents for a given patent application.

Prior-art search has particular challenges compared to typical search operations \cite{xue2009automatic}. Firstly, the patent documents are generally long and use  very technical language. Secondly, the documents are prepared to show the novelty of the application, instead of focusing on the similarities with the existing ones. Thirdly, it is a recall-oriented retrieval task as we have to find all relevant patents to detect the novelty of a patent application. 

Prior work shows that BERT \cite{bert} based models achieve state-of-the-art results in various Natural Language Processing (NLP) tasks. Therefore, in order to find relevant patents for a given patent, we can fine-tune a BERT \cite{bert} model to directly predict the  pairwise patent similarities.  However, this approach  has two main shortcomings. Firstly, BERT models are capable of processing only 512 ``tokens",%\footnote{Technically, three tokens are flag tokens. Therefore, the number of available tokens is 509.}, 
which corresponds to roughly 400 words on an average text. However, patents are generally much longer than 400 words and we have to provide two patent documents to calculate their similarity, reducing the number of tokens we can use for each patent. Therefore, this approach would force us to ignore many parts of patent documents. Secondly,  given that we have millions of patents,  predicting similarity scores using a fine-tuned BERT model for all patents for a given query patent would be excessively slow. 

In this paper,  we develop a novel method for overcoming the  shortcomings discussed above. In particular, we represent each patent using SciBERT \cite{scibert} allowing us to capture technical language used in patent documents. %and vectors provided by Google, separately. 
Next, we fine-tune SciBERT model based on Triplet Networks approach \cite{tripletnet}. This allows us to derive a fixed vector for each patent document and apply efficient vector computations. In query time, we rank patents based on their  cosine similarity to the query patent. % and the indexed patents using their SciBERT and Google's vector based representations, separately. Finally, we  rank patents based on their average similarity score.  
In our experiments with 1.8M patents, we show that our proposed method outperforms baseline methods.

%In this paper, we develop a   prior-search retrieval method. In particular, we represent each patent in two separate ways: 1) using  contextual embeddings of SciBERT\cite{scibert} and 2) patent vectors released by Google. 
%Subsequently, we train them separately using Triplet Networks \cite{tripletnet} approach. In query time, we calculate cosine similarity between the query and indexed patents using two different representations and rank them based on the average cosine similarity score. 

\section{Proposed Approach}

In this section, we explain the details of representing patents with SciBERT and  Triplet Network based fine-tuning. 
%since the system can accept only 256 tokens (roughly 200 words)\footnote{We assume that the number of tokens is divided equally for each patent in the pair.} for each patent, there is a significance loss in precision. Second, the BERT model should be called by the system \emph{N}$\times$\emph{(N-1)/2} times, where \emph{N} is the number of patents. Given that we have millions of patents, the implementation would take years to execute even with the current state-of-art computers.

% Deep-learning-based language models are useful to catch syntactic and semantic information within a text. In this project, 
%We convert each patent text into a numerical vector using state-of-the-art language models and calculate pairwise patent similarities to rank them. 
%We use text based patent representations to calculate a continuous\todo{bundan kastını anlamadım} similarity metric for patents. We use description and claims sections of patents. %which exist in most of the patents that are filed after 1976. 
\subsection{Patent Representation}
%We represent each patent using two vectors separately. Now we explain these representations. %use two different methods to represent patents.

%\subsubsection{BERT based representation}
%Bidirectional Encoder Representations from Transformers (BERT) \cite{bert} models achieve state-of-the-art results in various Natural Language Processing, suggesting that %the model 
%As the language model, we use Bidirectional Encoder Representations from Transformers (BERT) \cite{bert}, which %, which is developed by Google, 
%to convert patent texts into vectors. 
%achieves state-of-the-art results in various Natural Language Processing (NLP) tasks. One of the advantages of the BERT is that it processes words in relation to all other words in a sentence, rather than one-by-one or in a fixed-sized sliding window approach. Therefore, %Because of this, 
%the model is capable of understanding the context and intent behind a text. 
%BERT is originally trained using Wikipedia and BooksCorpus texts for two objectives to understand the language. 
%First one is called masked language modelling task in which 15\% of the words are replaced with a special token and model tries to find the original word(s). Latter one is called next sentence prediction task in which model tries to guess the next sentence for a given sentence. This two training phases are called pre-training which needs vast amount of computational power but it needs to be done once. Pre-trained models are released and then can be fine-tuned for specific task with much less data. 
 BERT models are successful at catching the semantics of texts. However, the language of patent documents might include many technical terminologies while BERT is pre-trained using Wikipedia articles and BooksCorpus. Therefore, %instead of using the original BERT pre-trained models, 
we exploit SciBERT \cite{scibert}, which is pre-trained on  large multi-domain corpus of scientific publications, instead of using the original BERT. 

%We can use BERT models to derive fixed-size embeddings for each patent text \cite{sentbert}. However,  BERT models are capable of processing only 512 ``tokens"\footnote{Technically, three tokens are flag tokens. Therefore, the number of available tokens is 509.}, which corresponds to roughly 400 words on an average text. On the other hand, 
Patent documents are generally much longer than BERT based models can process. We could truncate patent documents to meet the limits of BERT. However, it would mean ignoring many parts of patent documents that might be useful for our search task. Therefore, in order to capture the semantics of patent documents,  we create separate embeddings for the description ($v_d$) and claims ($v_c$) part of each patent.  For descriptions longer than 400 words, we use TextRank \cite{textrank} automatic summarization tool to reduce the text length to 400 words. However, for the claims part of patents, we do not use the text summarization but truncate the parts that exceed BERT's token limit. This is because the first claim of patents is generally the main innovative part of the patents while the other claims are less important ones. 
Subsequently,  we concatenate the vectors for the description and claim parts to form a single embedding for each patent and normalize them to have a unit norm. In order to give more emphasis to the description part of the patents than their claims,  
we multiply each element of $v_d$ by $\sqrt{0.8}$ and multiply each element of $v_c$ with $\sqrt{0.2}$. The parameters are selected arbitrarily. Note that because of the vector multiplication in cosine similarity calculation,  the relative weights used for description and claims parts will be 8:2. 

%SciBERT creates a vector with 768 elements for each token in the text. Since number of tokens can vary for each patent, we use a special token called ``CLS" to derive the  fixed-size embedding. Since SciBERT  can process only 512 tokens at one time, we create separate embeddings for description ($v_d$) and claims ($v_c$) part of each patent.  \newline

%Since deep-learning models require a high computational power, BERT is pre-trained using Wikipedia and BooksCorpus texts. Then, this pre-trained model is fine-tuned for a specific NLP task using an additional deep learning layer with a labeled data. In the fine-tuning process, we take the following steps to adapt the BERT model for the patent-similarity task. 

%\item BERT is capable of processing only 512 ``tokens"\footnote{Technically, three tokens are flag tokens. Therefore, the number of available tokens is 509.}, which corresponds to 400 words on an average text. Since patents are usually longer than 400 words, we use the description of patent texts to train the model. For descriptions longer than 400 words, we use TextRank \cite{textrank} automatic summarization tool to reduce the text size to 400 words. \newline
%\subsubsection{Google's Patent Vectors} Google provides numerical vectors for each patent\footnote{\url{https://console.cloud.google.com/marketplace/details/google\_patents\_public\_datasets/google-patents-research-data}}, enabling to conduct any textual analysis on patent documents. Separately from our BERT based representation, we also use vectors provided by Google.

\begin{table*}[!htb]
		\begin{tabular}{|l|c|c|c|c|}
			\hline
            \textbf{Ranking Method} & \textbf{Average Precision }      & \textbf{Recall@100 }& \textbf{Recall@500 }& \textbf{Recall@1000} \\ \hline
						Lucene with TF-IDF                              & 0.0548    & 0.2178    & 0.3642    & 0.4364     \\ \hline
			Lucene with BM25                                & 0.0469    & 0.1800      & 0.3083    & 0.3743     \\ \hline
%			Lucene with QL with Jelinek-Mercer                       & 0.0660     & 0.2521    & 0.4073    & 0.4816     \\ \hline \hline
			
%			Triple Network with Google Patent Vectors                              & 0.0601    & 0.2032    & 0.3780    & 0.4698     \\ \hline
	%		Bert\_Description                   & 0.0639    & 0.2102    & 0.3760    & 0.4637     \\ \hline
	%		Bert\_Claims                        & 0.0468    & 0.1713    & 0.3193    & 0.4002     \\ \hline
			%Bert\_Description\_Claims
			%Triple Network with SciBert Based Vectors 
			Our Approach & \textbf{0.0675}    & \textbf{0.2233}    & \textbf{0.3934}    & \textbf{0.4821}     \\ \hline
%		    Triple Network with both vectors   & \textbf{0.0797}    & \textbf{0.2605}    & \textbf{0.4518}    & \textbf{0.5444}     \\ \hline
		\end{tabular}
	\caption{Comparison of our approach with baseline methods. The best performing score for each metric is written in \textbf{bold}.}
	\label{tab:table-1}
\end{table*} 

\subsection{Fine-Tuning via Triplet Networks}
 We fine-tune SciBERT using Triplet Networks approach \cite{tripletnet} which allows us to derive fixed-size embeddings for each patent, and thereby,  apply efficient vector operations to calculate the similarity between patents. 
 In the Triplet Network approach, we have to provide positive and negative samples for each patent such that the model can learn the semantic differences between relevant and not relevant patents.
 In particular,  we construct 3 embeddings for each patent based on i) an anchor (i.e., the patent itself) patent (\textit{a}), ii) a positive (i.e., relevant) patent (\textit{p}),  and iii) a negative (i.e., not relevant) patent (\textit{n}). % (\textit{n}), which is a patent that does not cite the patent.
  We calculate triplet objective loss  as follows:

$$max(CosineDistance(v_a,v_p)- CosineDistance(v_a,v_n)+\epsilon,0)$$

\noindent
where $v_a$, $v_p$, and $v_n$ are the embeddings  for \textit{a}, \textit{p}, and \textit{n}, respectively. $\epsilon$ is a margin ensuring that $v_p$ is at least $\epsilon$ closer to $v_a$, than $v_n$.   

Obviously, the training data and the label distribution directly affect supervised models' performance.  
 Therefore, we take the following steps to select the patents given as positive and negative samples. 
%Since there are problems with the citation data and some cited-patents are in fact non-relevant, 
\begin{itemize}[leftmargin=0.1in]
\item We select `positive' texts from the cited patents which have a similarity score of higher than 0.6 according to vectors provided by Google\footnote{\url{https://console.cloud.google.com/marketplace/details/google\_patents\_public\_datasets/google-patents-research-data}}. %\todo{geri kalan yüzde 40ı ?}
\item  We select 20\% of the negatives from the not-cited patents which are from the  Cooperative Patent Classification (CPC) group of the anchor patent. 
%hav the first six digits in their Cooperative Patent Classification are same with the anchor patent's. are not cited by the anchor patent but have the same first six digits in their Cooperative Patent Classification (CPC) code. 
Therefore, the model can learn textual properties of patents that are on a similar topic but not as close as the positive ones.
\item  We select 20\% of the negatives from the patents which are not cited by the anchor patent but cited by the patents that it cites. This process allows us to train the models with negative samples that are not semantically far from the anchor patent.  %r but also not close enough to be cited. 

\item The remaining 60\% of the negatives are randomly selected from the patents which are not cited by the anchor patent and have a similarity score of less than 0.6 based on Google's vectors. Therefore, the model can learn the textual properties of patents that are distinctively different from the anchor.

\end{itemize}

%\subsection{Patent Retrieval Process}
%For a given patent document as a query, we represent it in two different vectors, one is based SciBERT  and the other one is based on Google's vector. Next, we calculate cosine similarity scores for each indexed patent and vector representation. Subsequently, we calculate average cosine similarity for each patent vector and rank them accordingly. 

% BERT can be used to calculate pairwise patent similarities in two ways. In this project, we chose the second approach.
%\begin{itemize}
% In the first approach, for any two patents that are passed to the BERT, the model directly provides a pairwise similarity score. Even though this approach has some advantages in specific tasks, it has two shortcomings for the purpose of this project. First, since the system can accept only 256 tokens (roughly 200 words)\footnote{We assume that the number of tokens is divided equally for each patent in the pair.} for each patent, there is a significance loss in precision. Second, the BERT model should be called by the system \emph{N}$\times$\emph{(N-1)/2} times, where \emph{N} is the number of patents. Given that we have millions of patents, the implementation would take years to execute even with the current state-of-art computers.

\section{Experiments} \label{sec_exp}

%\subsection{Experimental Setup}

 We randomly select 2 million patents granted after 1980. Among these patents, 1,817,504 of them have a title, abstract, description, and claims sections. From this sample, we randomly select 5,000 patents for testing, and others are used in training. Following prior work \cite{xue2009automatic}, we consider cited patents as relevant ones and not-cited ones as not-relevant. %---these patents are not included in any training data.

%We first train the model with 2 million examples (i.e., patent triplets) with $\epsilon=0.5$, then train with additional 2 million examples with $\epsilon=0.05$.
We train the model with four million examples (i.e., patent triplets). 
 We use patents %with at that 
which have at least five backward and forward citations in total, %cited patents and patents that cite the respective patent), 
as anchors in the training set. We train the model using 4 Nvidia Titan RTX GPUs with a batch size of 8, using Adam optimizer with a learning rate of $3e^{-6}$ with linear learning rate warm-up over 10\% of the training data for 1 epoch.

We compare our model against   BM25 and TF-IDF ranking functions that  Lucene\footnote{\url{https://lucene.apache.org/core/}}  provides.
%BM25, TF-IDF, and  Query Likelihood (QL) with Jelinek-Mercer smoothing. We also evaluate performance of our model when we use only Google's patent vectors and only SciBERT based vectors. 
The results are shown in \textbf{Table~\ref{tab:table-1}}. %Our observations are as follows. Firstly, our approach in which we combine Google vectors and SciBERT based vectors outperform all other methods in all cases. Secondly, all triplet network based methods outperform TF-IDF and BM25 based ranking. However, QL with Jelinek-Mercer smoothing achieve higher recall@100 and recall@500 scores than the methods when we use only Google's vectors or only SciBERT based vectors. Lastly, our SciBERT based vectors yields higher performance than Google's vectors in all cases.
We observe that  our approach outperforms Lucene's methods based on all four metrics, suggesting that our proposed method can be an effective solution for the prior-art search problem. 

\section{Conclusion}
In this paper, we propose a novel method to represent patent documents by fine-tuning SciBERT with  Triplet Network approach. We show that our proposed method outperforms baseline methods in our experiments. In the future, we plan to extend our work in several directions. Firstly, we plan to use other variants of BERT  pre-trained with different types of documents, e.g., PatentBERT. In addition, we plan to investigate which parts of patent documents are more important for the prior-art search task and how to best summarize them. Furthermore, we will investigate using BERT variants that have higher token limits. Finally, we believe that our model should be evaluated in various test collections and compared against other baseline methods.  

\section*{Acknowledgment} This study was funded by the Scientific and Technological Research Council of Turkey (TUBITAK) ARDEB 1001 Grant No 119K986.  The statements made herein are solely the responsibility of the authors. 

%the following baselines.
%\begin{itemize}
%    \item Google Patent Vectors: Google provides numerical vectors for each patent. We use these vectors to calculate Cosine similarity between patents. The patents are ranked accordingly.
%    \item Our Model + Google Vectors: We take the average of similarity scores based on our model and Google Patent Vectors.
 %   \item Lucene: We use to rank similar patents. Lucene provides several ranking functions. We use TF-IDF, BM25, and Query Likelihood (QL) with Jelinek-Mercer smoothing. 
    %Term Frequency-Inverse document Frequency (TF-IDF) is a method which statistically measures how import a word is to a document in a collection. Term frequency measures how frequently a word occurs in a document. Inver
%\end{itemize}

%\section{Conclusion}

\bibliographystyle{ACM-Reference-Format}
\bibliography{sample-base}
\end{document}